\documentclass[letter]{aa}
\usepackage[varg]{txfonts}
\usepackage{graphicx}
\usepackage{morefloats}
\usepackage{natbib}
\bibpunct{(}{)}{;}{a}{}{,}

\newcommand{\lsio}{\object{LS I +61 303}}
\newcommand{\lsi}{LS~I~+61~303}

\begin{document}

\titlerunning{Evidence of coupling between the thermal and nonthermal emission in \lsi{}}
\title{Evidence of coupling between the thermal and nonthermal emission in the gamma-ray binary \lsi{}
\thanks{Tables 1 and 2 are only available in electronic form at the CDS via anonymous ftp to {\tt cdsarc.u-strasbg.fr} ({\tt 130.79.128.5}) or via {\tt http://cdsweb.u-strasbg.fr/cgi-bin/qcat?J/A+A/}}
}

\author{X. Paredes-Fortuny\inst{1}
\and M. Rib\'o\inst{1,}\thanks{Serra H\'unter Fellow.}
\and V. Bosch-Ramon\inst{1}
\and J. Casares\inst{2,3}
\and O. Fors\inst{4,1}
\and J. N\'u\~nez\inst{1,5}}

\institute{Departament d'Astronomia i Meteorologia, Institut de Ci\`ences del Cosmos, Universitat de Barcelona, IEEC-UB, Mart\'{\i} i Franqu\`es 1, E-08028 Barcelona, Spain, e-mail: xparedes@am.ub.es
\and
Instituto de Astrof\'{\i}sica de Canarias, E-38205 La Laguna, Santa Cruz de Tenerife, Spain
\and
Departamento de Astrof\'{\i}sica, Universidad de La Laguna,
E-38206 La Laguna, Santa Cruz de Tenerife, Spain
\and
Department of Physics and Astronomy, University of North Carolina at Chapel Hill, Chapel
Hill, NC 27599-3255, USA
\and
Observatori Fabra, Reial Acad\`emia de Ci\`encies i Arts de Barcelona, 
Rambla dels Estudis, 115, 08002 Barcelona, Spain
}

\date{Received - / Accepted -}

\abstract 
{
The gamma-ray binary \lsio{} is composed of a Be star and a compact companion orbiting in an eccentric orbit. Variable flux modulated with the orbital period of $\sim$26.5~d has been detected from radio to very high-energy gamma rays. In addition, the system presents a superorbital variability of the phase and amplitude of the radio outbursts with a period of $\sim$4.6~yr. We present optical photometric observations of \lsi{} spanning $\sim$1.5~yr and contemporaneous H$\alpha$ equivalent width ($EW_{\rm H{\alpha}}$) data. The optical photometry shows, for the first time, that the known orbital modulation suffers a positive orbital phase shift and an increase in flux for data obtained 1-yr apart. This behavior is similar to that already known at radio wavelengths, indicating that the optical flux follows the superorbital variability as well. The orbital modulation of the $EW_{\rm H{\alpha}}$ presents the already known superorbital flux variability but shows, also for the first time, a positive orbital phase shift. In addition, the optical photometry exhibits a lag of $\sim$0.1--0.2 in orbital phase with respect to the $EW_{\rm H{\alpha}}$ measurements at similar superorbital phases, and presents a lag of $\sim$0.1 and $\sim$0.3 orbital phases with respect noncontemperaneous radio and X-ray outbursts, respectively. The phase shifts detected in the orbital modulation of thermal indicators, such as the optical flux and the $EW_{\rm H{\alpha}}$, are in line with the observed behavior for nonthermal indicators, such as X-ray or radio emission. This shows that there is a strong coupling between the thermal and nonthermal emission processes in the gamma-ray binary \lsi{}.
The orbital phase lag between the optical flux and the $EW_{\rm H{\alpha}}$ is naturally explained considering different emitting regions in the circumstellar disk, whereas the secular evolution might be caused by the presence of a moving one-armed spiral density wave in the disk.
}

\keywords{X-rays: binaries -- X-rays: individual: \lsi{} -- binaries: close -- 
stars: emission-line, Be -- gamma rays: stars }
    
\maketitle

\section{Introduction}
\label{introduction}

\lsio{} is one of the five gamma-ray binaries currently known (e.g., \citealt{Dubus2013, Paredes2013}). It is composed of an optical star ($V$$\sim$10.7) with spectral type B0~Ve, therefore presenting a circumstellar disk, and a compact companion that is probably a pulsar orbiting in an eccentric orbit with $e$ in the range 0.54--72 (see \citealt{Hutchings1981, Paredes1986, Casares2005, Dhawan2006, Aragona2009}). The distance to the source is estimated to be $\sim$2 kpc based on \ion{H}{i} measurements (\citealt{Frail1991}). The binary system has an orbital period of $26.4960\pm0.0028$~d (\citealt{Gregory2002}), and variable flux modulated with the orbital period has been detected from radio to very high-energy gamma rays (e.g., \citealt{Taylor1982, Mendelson1989, Paredes1994, Paredes1997, Abdo2009, Albert2009}). In addition to the orbital modulation, \lsi{} also exhibits a superorbital modulation of radio outbursts in $\sim$4.6~yr (first detected by \citealt{Paredes1987} and \citealt{Gregory1989}). \cite{Gregory2002} found a superorbital period of $1667\pm8$~d in the amplitude and in the orbital phase of the radio outbursts, leading to a drift of the outburst maxima from orbital phase $\sim$0.4 to $\sim$0.9. For reference, the periastron passage takes place at orbital phase 0.23--0.28 \citep{Casares2005, Aragona2009}. The zero orbital and superorbital phases are set at JD~2,443,366.775 \citep{Gregory2002}.

\cite{Zamanov2013} have recently detected the $\sim$4.6~yr superorbital period in the equivalent width of the ${\rm H{\alpha}}$ emission line, confirming the earlier evidences reported in \cite{Zamanov1999}. This suggests that the superorbital variability is related to periodic changes in the mass-loss rate of the Be star and/or variations in the circumstellar disk (see \citealt{Massi2014} and references therein for alternative interpretations). \cite{Li2012} discovered the superorbital modulation of \lsi{} in X-rays (3--30~keV), with a period compatible with that found by \cite{Gregory2002} from radio measurements. \cite{Chernyakova2012} also reported the superorbital modulation of the X-ray emission, and using contemporaneous X-ray and radio observations showed that radio outbursts lag the X-ray outbursts by $\sim$0.2 in orbital phase along the entire superorbital cycle. 
\cite{Ackermann2013} found a sinusoidal variability in the $>$100~MeV $\gamma$-ray flux consistent with the radio superorbital period.
Finally, \cite{Zaitseva2003} found long-term variability in the mean $V$-band magnitude of 0.07~mag,
but no search for superorbital variability was conducted.

\begin{figure*}[!ht]
\begin{center}
\resizebox{0.59\hsize}{!}{\includegraphics{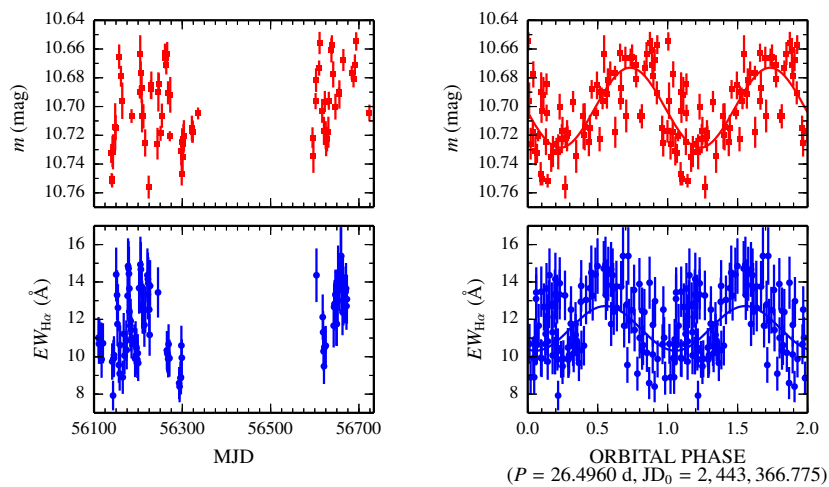}}
\vspace{-2mm}
\caption{
Optical photometric lightcurve (top, red) and $EW_{\rm H{\alpha}}$ (bottom, blue) of \lsi{} plotted as a function of MJD (left) and folded with the orbital phase (right).
Error bars represent 1-$\sigma$ uncertainties. 
The solid curves in the right-hand panels represent sinusoidal fits to the orbital variability, whose parameters can be found in Table~\ref{tab}.
Two cycles are displayed for clarity.}
\label{total}
\end{center}
\end{figure*}

In this work, we present optical photometric observations of \lsi{} spanning $\sim$1.5~yr, and contemporaneous H$\alpha$ equivalent width ($EW_{\rm H{\alpha}}$) observations. We report the discovery of an orbital phase shift and a variation in the orbitally modulated flux, very similar to the well-known superorbital behavior of the orbital modulation in radio. In addition, we also detect a phase shift in the orbital modulation of the $EW_{\rm H{\alpha}}$, to our knowledge reported here for the first time. We discuss the obtained results in the context of the radio and X-ray superorbital variability, and conclude that there is a strong coupling between the thermal and nonthermal emission in \lsi{}.

\section{Observations and data reduction}
\label{observations}

We performed optical photometric observations of \lsi{} with the robotic Telescope Fabra-ROA Montsec (TFRM; see \citealt{Fors2013}). The TFRM is installed at the Observatori Astron\`omic del Montsec (Lleida, Spain). The main specifications are: corrector plate of 0.5~m aperture and 0.78~m primary mirror, refurbished Baker-Nunn Camera for routine CCD robotic observations, focal ratio f/0.96, $\rm{4.4^{\circ}\times 4.4^{\circ}}$ field of view with a pixel scale of $3.9{^\prime}{^\prime}$/pixel, passband filter SCHOTT GG475 (${\lambda}>$ 475~nm.), and custom CCD based on FLI ProLine 16803 with quantum efficiency of 60\% at 550~nm (\citealt{Fors2013}).

The observations span from 2012 July 31 to 2014 March 7 (2 seasons) with 71 nights of good data (preliminary results on season 1 data were presented in \citealt{Paredes-Fortuny2014}). We observed the target around 20 times per night with exposures of 5--10~s, as a compromise between good signal to noise ratio and avoiding the nonlinear CCD regime.

We conducted the data reduction and analysis using a pipeline developed in Python following these steps: standard calibration of the images using IRAF\footnote{IRAF is distributed by NOAO, which is operated by AURA, under cooperative agreement with NSF.} (implemented through PyRAF\footnote{PyRAF is a product of the Space Telescope Science Institute, which is operated by AURA for NASA.}),
aperture photometry using the PHOT package from IRAF (PyRAF) with an aperture radius of 7.5 pixels, and correction of the lightcurves using a weighted average differential magnitude correction method based on \cite{Broeg2005}. The lightcurve of \lsi{} has been corrected using 145 weighted reference stars up to 4~mag fainter than the target and as close as possible to it ($<0.25\degr$). Finally, we averaged the
corrected magnitudes of the target on a nightly basis, and we used an artificial offset to a mean magnitude of 10.7~mag. The nightly photometric uncertainties are estimated as the standard deviation of the magnitudes of the target obtained from the individual images for each night. The obtained values are typically in the 5--10~mmag range.

We also considered contemporaneous $EW_{\rm H{\alpha}}$ data of \lsi{} obtained using FRODOspec on the robotic 2.0-m Liverpool telescope at the Observatorio del Roque de Los Muchachos (La Palma, Spain). The data span from 2012 July 2 to 2014 January 15 with 104 measurements, with one 600~s exposure per night. The data have been obtained with the same setup, and reduced in the same way, as in \cite{Casares2010}. We assumed that the uncertainties of the $EW_{\rm H{\alpha}}$ are at the 10\% level. The adopted convention is positive $EW$ for emission.

The data (Tables 1 and 2) are available online at the CDS.

\section{Results}
\label{results}

\begin{figure*}[!ht]
\begin{center}
\resizebox{0.59\hsize}{!}{\includegraphics{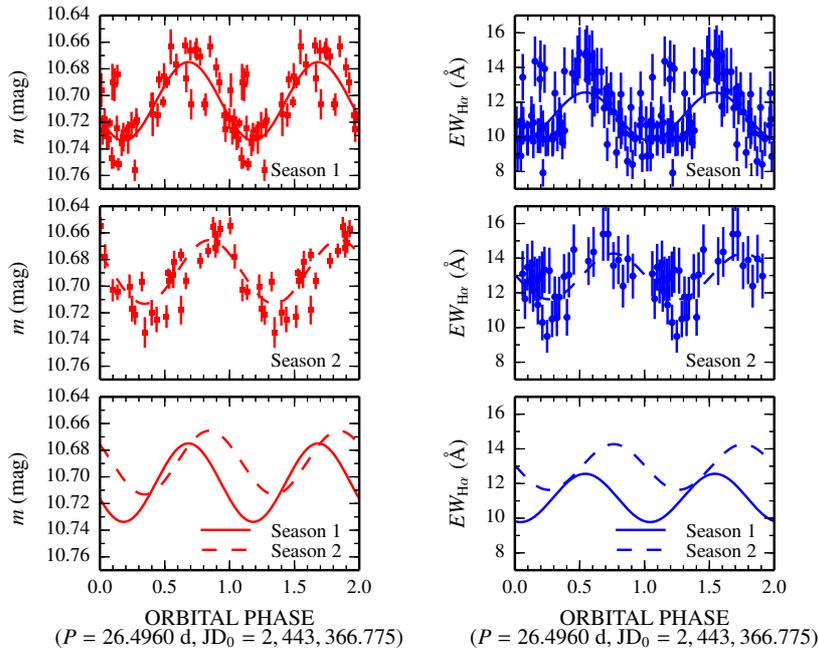}}
\vspace{-2.5mm}
\caption{
Optical photometric lightcurve (left, red) and $EW_{\rm H{\alpha}}$ (right, blue) of \lsi{} folded with the orbital phase for the first (top) and second (center) seasons, and their sinusoidal fits to the orbital variability (bottom; see Table~\ref{tab}). 
Error bars represent 1-$\sigma$ uncertainties.
Two cycles are displayed for clarity.
The shifts in the orbital modulations between both seasons are clear.
}
\label{seasons}
\end{center}
\end{figure*}

The optical photometric lightcurve and $EW_{\rm H{\alpha}}$ of \lsi{} as a function of MJD and folded with the orbital phase are shown in Fig.~\ref{total}. As can be seen, the data cover two different seasons (centered in autumn) and show the already known orbital variability (see, e.g., \citealt{Zamanov2014}). The folded lightcurves for the first and second season and their sinusoidal fits are shown in Fig.~\ref{seasons}. The fitted sinusoidal parameters (amplitude, zero phase, and mean value) for the different data sets, considering the data uncertainties, are quoted in Table~\ref{tab}. The 1-$\sigma$ uncertanties in the fitted parameters quoted in Table~\ref{tab} are estimated from the covariance matrix.

\begin{table*}[!ht]
\renewcommand\thetable{3} 
\caption{Fitted sinusoidal parameters to the orbital variability of the optical photometry and the $EW_{\rm H{\alpha}}$ of \lsi{} for the total data set (Fig.~\ref{total}) and for the two different observational seasons (Fig.~\ref{seasons}). The parameters are those of the functions: optical $m=-A \cos \left( 2\pi \left[ \phi-\phi_0\right] \right)+C$ and $EW_{\rm H{\alpha}}= A \cos \left( 2\pi \left[ \phi-\phi_0\right] \right)+C$. Phases were computed using an orbital period of 26.4960~d and phase zero at JD~2,443,366.775.}
\label{tab}
\centering
\begin{tabular}{lc@{~~~}c@{~~~}c@{~~~~}c@{~~~~}c@{~~~}c@{~~~}c@{~~~}c}
\hline\hline
Data set & MJD$^{\rm optical}$ & $A^{\rm optical}$ (mmag) & $\phi_0^{\rm optical}$ & $C^{\rm optical}$ (mag) & MJD$^{EW_{\rm H{\alpha}}}$ & $A^{EW_{\rm H{\alpha}}}$ ($\AA$) & $\phi_0^{EW_{\rm H{\alpha}}}$ & $C^{EW_{\rm H{\alpha}}}$ ($\AA$) \\
\hline
Total    & 56139--56723 & $28\pm3$ & $0.73\pm0.02$ & $10.701\pm0.002$ & 56110--56672 & $1.2\pm0.2$ & $0.56\pm0.03$ & $11.5\pm0.2$ \\ 
Season 1 & 56139--56335 & $29\pm3$ & $0.68\pm0.02$ & $10.704\pm0.003$ & 56110--56298 & $1.4\pm0.3$ & $0.54\pm0.03$ & $11.2\pm0.2$ \\ 
Season 2 & 56594--56723 & $24\pm4$ & $0.85\pm0.02$ & $10.689\pm0.003$ & 56603--56672 & $1.3\pm0.3$ & $0.76\pm0.05$ & $13.0\pm0.3$ \\ 
\hline
\end{tabular}
\end{table*}

The fits to the orbital variability of the optical photometry reveal, for the first time, a positive orbital phase shift of $0.16\pm0.03$ at 5.2~$\sigma$ confidence level (c.l.) and an increase in the average optical flux of $15\pm4$~mmag at 4.2~$\sigma$ c.l. between the two observational seasons 1-yr apart. These trends are reminiscent of the superorbital trends found at radio and X-ray wavelengths, and indicate that the optical flux follows the superorbital variability as well. We note that the folded lightcurves of the two individual seasons (see Fig.~\ref{seasons}-left) show less scatter than the folded lightcurve including all the data (see Fig.~\ref{total}), which span $\sim$0.35 superorbital phases.

Similarly, the fits to the orbital variability of $EW_{\rm H{\alpha}}$ present for the first time a positive orbital phase shift of $0.22\pm0.05$ at 4.2~$\sigma$ c.l. Further, we confirm the previously claimed superorbital modulation of the mean $EW_{\rm H{\alpha}}$, with a variation of $1.8\pm0.3$~\AA\ at 5.7~$\sigma$ c.l., between our two observational seasons\footnote{Despite the relatively poor sampling at equivalent superorbital phases, the \cite{Zamanov2013} data reveal a compatible behavior.}. In addition, the orbital modulation of the optical photometry exhibits a lag of $\sim$0.1--0.2 in orbital phase with respect to the orbital modulation of  $EW_{\rm H{\alpha}}$ at similar superorbital phases (see Fig.~\ref{seasons} and Table~\ref{tab}).

\begin{figure*}
\begin{center}
\resizebox{0.91\hsize}{!}{\includegraphics{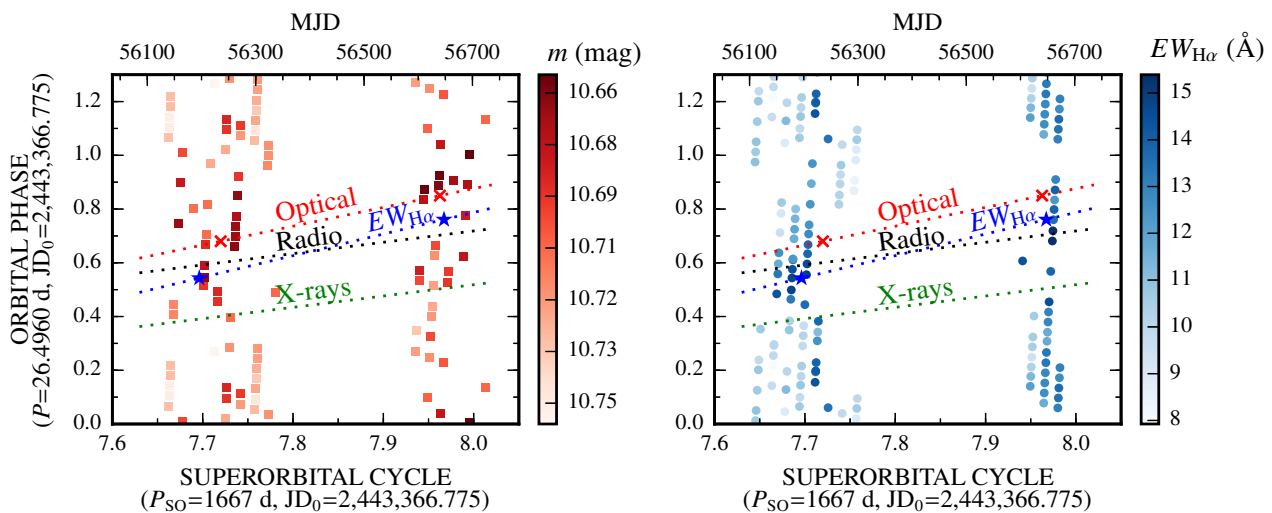}}
\vspace{-3mm}
\caption{
Color maps of the optical photometry (left) and $EW_{\rm H{\alpha}}$ (right) from \lsi{} as a function of the orbital phase and superorbital cycle. The red crosses and blue stars correspond to the phases of the maxima of the sinusoidal fits to the orbital variability of the optical photometry and $EW_{\rm H{\alpha}}$, respectively, for the first and second seasons (see Table~\ref{tab}). The dotted color lines represent the orbital phase drift of the emission peaks caused by the superorbital variability for the optical photometry (red), $EW_{\rm H{\alpha}}$ (blue), radio (black), and X-rays (green). The radio and X-ray drifts are taken from Fig.~3 of \cite{Chernyakova2012} using data of the previous superorbital cycle.
}
\label{phases}
\end{center}
\end{figure*}

To better display the superorbital variability, color maps of the optical
photometry and $EW_{\rm H{\alpha}}$ as a function of the orbital phase and
superorbital cycle are shown in Fig.~\ref{phases}. The dotted red and blue lines represent the orbital phase drifts of the corresponding maxima along the
superorbital cycle for the contemporaneous optical photometry and $EW_{\rm
H{\alpha}}$ presented here (superorbital cycle 7). The contemporaneous radio
and X-ray fluxes of the previous superorbital cycle (6) from
\cite{Chernyakova2012} are shown as black and green dotted lines. The optical photometric observations show a lag of $\sim$0.1 in orbital phase with respect to the radio outburst, and $\sim$0.3 with respect to the X-ray outbursts for equivalent superorbital phases one cycle apart. The $EW_{\rm H{\alpha}}$ maxima occur at orbital phases similar to the radio outbursts for similar superorbital phases one cycle apart, while in the averaged data of \cite{Zamanov2013, Zamanov2014} the maxima occur during the rising of the radio flux density. The different slopes in Fig.~\ref{phases} might suggest a change in the behavior of the superorbital modulation between different superorbital cycles, although we caution that the results are still compatible at 2$\sigma$ c.l.

\section{Discussion}
\label{discussion}

The observations of \lsi{} reported here show, for the first time, orbital phase shifts in the maxima of the orbitally modulated optical flux and $EW_{\rm H{\alpha}}$ for different phases of the superorbital cycle (see Figs.~\ref{seasons} and \ref{phases} and Table~\ref{tab}). These results extend previously reported links between the H$\alpha$ emission and multiwavelength properties but obtained from the \emph{average} orbital modulation, i.e., without considering the superorbital variability (see, e.g., \citealt{Zamanov2014}). The results reported here are now discussed in the context of the known multiwavelength behavior of the orbital and superorbital variability of the source.

The orbital variability of the gamma-ray binary \lsi{} is wavelength dependent. After periastron at phase 0.23--0.28, there is first a maximum in X-rays (good tracer of the nonthermal emission), followed by a maximum in $EW_{\rm H{\alpha}}$ (tracer of the outer disk conditions) and in radio emission (produced mainly outside the binary system). Finally, a maximum in optical flux is observed, with a 65\% contribution from the Be star plus 35\% from the inner circumstellar disk \citep{Casares2005}. The circumstellar disk is likely to be perturbed and/or partially disrupted by tidal forces and the putative pulsar wind ram pressure. These would trigger significant changes in the structure of the Be disk, especially around periastron passage, reducing its emitting area and total optical emission. This is approximately what is observed in the data from season~1,
but the orbital phase shift in season~2 clearly shows that there is more than purely orbit-induced variability. In fact, the superorbital variability observed in $EW_{\rm H{\alpha}}$ has been associated with periodic changes on the Be star envelope and its circumstellar disk (e.g., \citealt{Zamanov1999}), which would trigger the superorbital variability observed at other wavelenghts. If this is the case, since the orbital modulation seen in X-rays or radio suffers a phase drift along the superorbital cycle, we should detect the same effect in optical flux and $EW_{\rm H{\alpha}}$, as reported here for the first time. Thus, it is clear that there is a strong empirical coupling between the thermal (optical) and the nonthermal (X-ray and radio) emission in the gamma-ray binary \lsi{} at both orbital and superorbital scales.

The orbitally modulated optical flux and $EW_{\rm H{\alpha}}$ present different behavior between them. 
The optical flux 
shows $\sim$0.06~mag modulation, representing $\sim$6\% in total flux or $\sim$16\% in disk flux (and probably projected area), while $EW_{\rm H{\alpha}}$ shows $\sim$30\% variability. This implies that external parts of the disk are more perturbed or disrupted than the inner parts of the disk, as one would expect if these perturbations (or disruptions) are due to the influence of the compact object as it approaches periastron (where the perturbations are caused by tidal forces and/or ram pressure). In addition, the $\sim$0.1--0.2 phase lag implies that the external parts are perturbed or disrupted before the internal parts, but they recover earlier as well, probably building up from material shocked before periastron passage. The inner disk would only recover close to, or even after, apastron. The secular evolution of this behavior along the superorbital cycle could be due to the presence of a moving one-armed spiral density wave in the disk, as suggested by \cite{Negueruela1998} for traditional Be/X-ray binaries.

\begin{acknowledgements}
The authors acknowledge support of the TFRM team for preparing and carrying out the optical photometric observations.
We thank useful discussions with Josep M. Paredes and Ignacio Negueruela.
We acknowledge support by the Spanish Ministerio de Econom\'{\i}a y Competitividad (MINECO) under grants 
AYA2013-47447-C3-1-P, 
AYA2010-18080,
and 
FPA2013-48381-C6-6-P.
This research has been supported by the Marie Curie Career Integration Grant
321520.
X.P.-F. acknowledges financial support from Universitat de Barcelona and Generalitat de Catalunya under grants APIF and FI (2014FI\_B 01017), respectively.
V.B-R. acknowledges financial support from MINECO and European Social Funds through a Ram\'on y Cajal fellowship.
\end{acknowledgements}

\bibliographystyle{aa}
\bibliography{bibliography.bib}

\end{document}